\documentclass{elsartchimie}

\usepackage{graphicx}
\usepackage{amssymb,amsbsy,amsmath}

\newcommand{\op}[1]{%
    \fontdimen12\textfont3=2pt\fontdimen12\scriptfont3=1.4pt%
    \!\null\mathop{\vphantom{#1}\smash{#1}}\limits_{\sim}\null\!}
\newcommand{\xref}[1]{\protect\ref{#1}}
\newcommand{\figref}[1]{Fig.~\protect\ref{#1}}

\newcommand {\mofe} {\{$\textrm{Mo}_{72}\textrm{Fe}_{30}$\}}

\begin{document}
\begin{frontmatter}

\title{Frustration-induced exotic properties of magnetic molecules}

\author{J\"urgen Schnack\corauthref{cor1}}
\address{Universit\"at Osnabr\"uck, Fachbereich Physik,
D-49069 Osnabr\"uck, Germany}
\corauth[cor1]{Tel: ++49 541 969-2695; fax: -12695; Email: jschnack@uos.de}

\begin{abstract}
  Geometric frustration of interacting spin systems is the
  driving force of a variety of fascinating phenomena in
  low-dimensional magnetism. In this contribution I will review
  recent results on frustration-induced effects in magnetic
  molecules, i.e. zero-dimensional magnetic systems, as well as
  in a recently synthesized frustrated molecule-based spin tube,
  i.e. a one-dimensional spin system.
\end{abstract}

\end{frontmatter}

\section{Introduction}
\label{sec-1}

Antiferromagnets with competing interactions are said to be
geometrically frustrated. As a result they exhibit a variety of
fascinating phenomena \cite{Gre:JMC01}. In this context the term
\emph{frustration} describes a situation where in the ground
state of the corresponding classical spin system not all
interactions can be saturated simultaneously. A typical picture
for such a situation is a triangle of antiferromagnetically
coupled spins, where classically the spins are not in the
typical up-down-up configuration, but assume a ground state that
is characterized by a relative angle of $120^\circ$ between
neighboring spins.  This special type of classical ground state
characterizes several frustrated spin systems, among them giant
Keplerate molecules \cite{MLS:CPC01,MTS:AC05,BKH:CC05}, the
triangular lattice antiferromagnet, and the kagome lattice
antiferromagnet.

Among the phenomena caused by frustration are magnetization
plateaus and jumps as well as unusual susceptibility minima, as
observed for example for the kagome lattice antiferromagnet
\cite{NKH:EPL04,SHS:PRL02}.  Some of these effects can also
occur in certain strongly frustrated magnetic molecules such as
the Keplerate \mofe \cite{SNS:PRL05} and in a cuboctahedral
molecule \cite{SSR:JMMM05} which has been already synthesized as
\{Cu$_{12}$La$_8$\}, but has not yet been magnetically
characterized \cite{BGG:JCSDT97}.  It has also soon be noticed
that the high degeneracy of magnetic levels at the saturation
field, where the unusual jumps occur, should also have an
enormous impact on the magnetocaloric effect.  An adiabatic
process should yield a massively enhanced cooling rate in the
vicinity of large magnetization jumps.  Such a behavior was
first anticipated for the classical version of the kagome
lattice antiferromagnet \cite{Zhi:PRB03} and subsequently
demonstrated for certain low-dimensional quantum spin systems
\cite{ZhH:JSM04}. In addition it could be shown that this effect
is rather stable against perturbations \cite{DeR:PRB04}. In the
following section some of these effects will be reviewed in more
detail.

Interlinking the zero-dimensional molecular units to
cluster-based networks is a highly interesting new route in
molecular magnetism, especially if the degree of magnetic
exchange between the cluster entities can be varied over a wide
range. Examples for such cluster-based networks are e.~g. given
by chains \cite{MDT:ACIE02} and square
lattices \cite{MSS:ACIE00,MSS:SSS00} of the magnetic Keplerate
molecule \{$\textrm{Mo}_{72}\textrm{Fe}_{30}$\}. These systems
show new combinations of physical properties that stem from both
molecular and bulk effects. As an example for a quasi
one-dimensional molecule-based frustrated spin system the
recently synthesized three-leg ladder compound
[(CuCl$_2$tachH)$_3$Cl]Cl$_2$ will be discussed
\cite{SKK:CC04,SNK:PRB04}.

Throughout the article the spin systems are modeled by an
isotropic Heisenberg Hamiltonian augmented with a Zeeman term,
i.e.,
\begin{eqnarray}
\label{E-2-1}
\op{H}
&=&
-
\sum_{u, v}\;
J_{uv}\,
\op{\vec{s}}(u) \cdot \op{\vec{s}}(v)
+
g \mu_B B \op{S}_z
\ .
\end{eqnarray}
$\op{\vec{s}}(u)$ are the individual spin operators at sites
$u$, $\op{\vec{S}}$ is the total spin operator, and $\op{S}_z$
its $z$-component along the homogeneous magnetic field axis.
$J_{uv}$ are the matrix elements of the symmetric coupling
matrix. In the following we will consider only antiferromagnetic
couplings which are characterized by negative values of
$J_{uv}$.

\section{Frustration effects in magnetic molecules}

For not too strong frustration, i.e. not too large deviation
from bipartiteness \cite{LSM:AP61,LiM:JMP62}, many spin systems
possess properties that are similar to or in accord with their
unfrustrated, i.e. bipartite counterparts. This is for instance
the case for antiferromagnetically coupled Heisenberg spin
rings, which -- if even-membered -- are covered by such famous
rigorous results as the sign rule of Marshall and Peierls
\cite{Mar:PRS55} and the theorems of Lieb, Schultz, and Mattis
\cite{LSM:AP61,LiM:JMP62}. A key quantity of interest is the
shift quantum number $k=0,\dots N-1$ associated with the cyclic
shift symmetry of the rings. The corresponding crystal momentum
is then $2\pi k/N$. For the aforementioned rings with even $N$
(bipartite) one can explain the shift quantum numbers for the
relative ground states in subspaces ${\mathcal H}(M)$ of total
magnetic quantum number $M$ \cite{LSM:AP61,LiM:JMP62,Mar:PRS55}.
In recent investigations we could numerically verify and prove
in some cases, that even for frustrated rings with odd $N$
astonishing regularities hold, and thus a generalized sign rule
can be established, which holds for chains of arbitrary sizes
\cite{BHS:PRB03}.

An antiferromagnet that can be decomposed into two sublattices
has as its lowest excitations the rotation of the N\'eel vector
as well as spin wave excitations \cite{And:PR52}. In finite size
systems these excitations are arranged in rotational (parabolic)
bands of energy eigenvalues as a function of total spin. Such a
behavior is most pronounced for bipartite, i.e. unfrustrated
systems \cite{ScL:PRB01,Wal:PRB02,WGC:PRL03}. It turns out that
the idea of rotational bands can be used to approximate the
low-lying energy spectrum of the Keplerate molecule \mofe\ which
due to the huge size of the Hilbert space practically cannot be
dealt with by other quantum methods \cite{SLM:EPL01}. Such an
approximation can explain the field dependence of the
magnetization at low temperatures \cite{SLM:EPL01}. It also
predicts a strong resonance in the inelastic neutron scattering
cross section at the separation between the two lowest bands
which was indeed discovered in a recent experiment using
inelastic neutron scattering \cite{GNZ:05}. 

\begin{figure}[ht!]
\centering
\centerline{\includegraphics[clip,width=50mm]{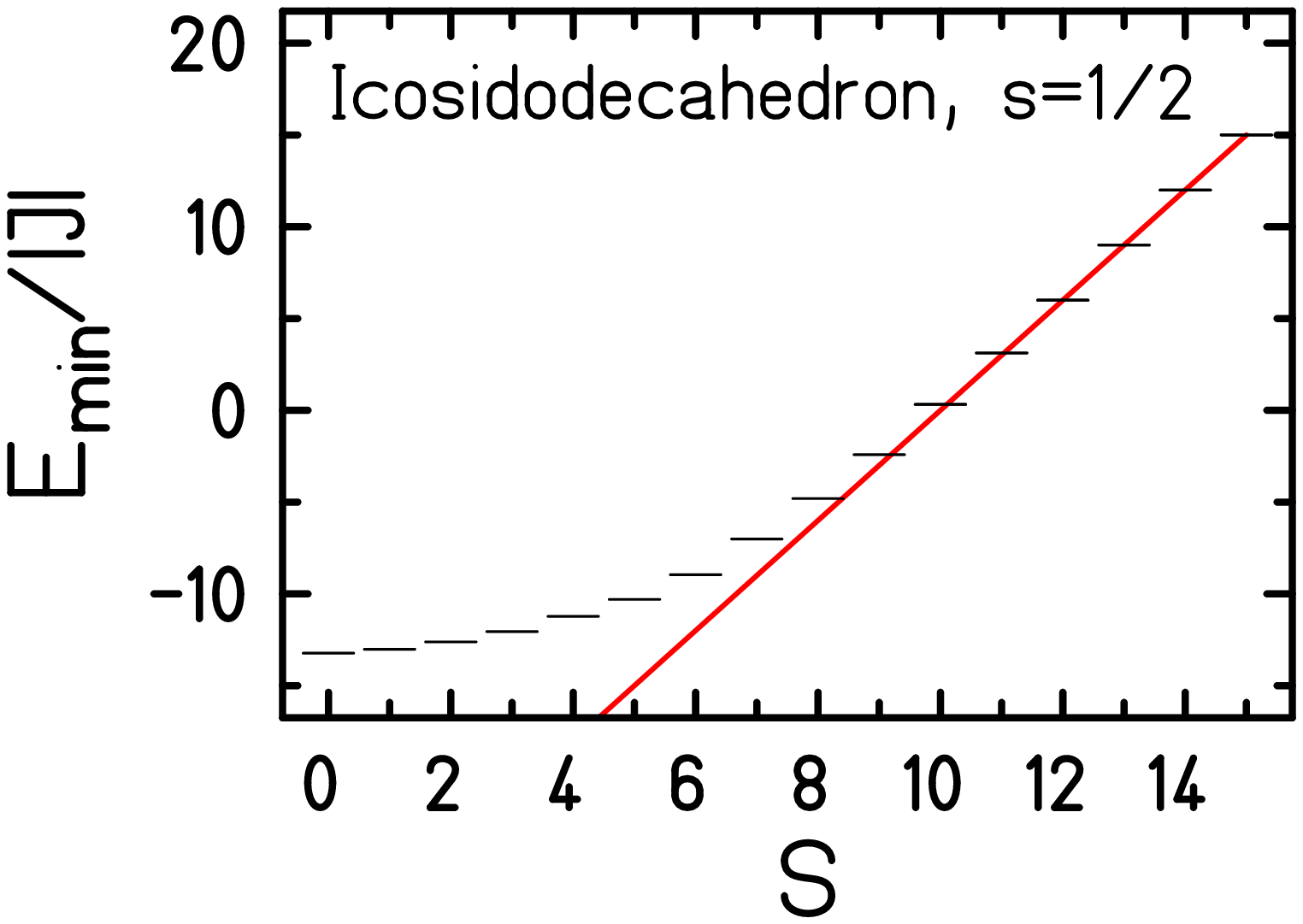}}

\vspace*{5mm}

\centerline{\includegraphics[clip,width=50mm]{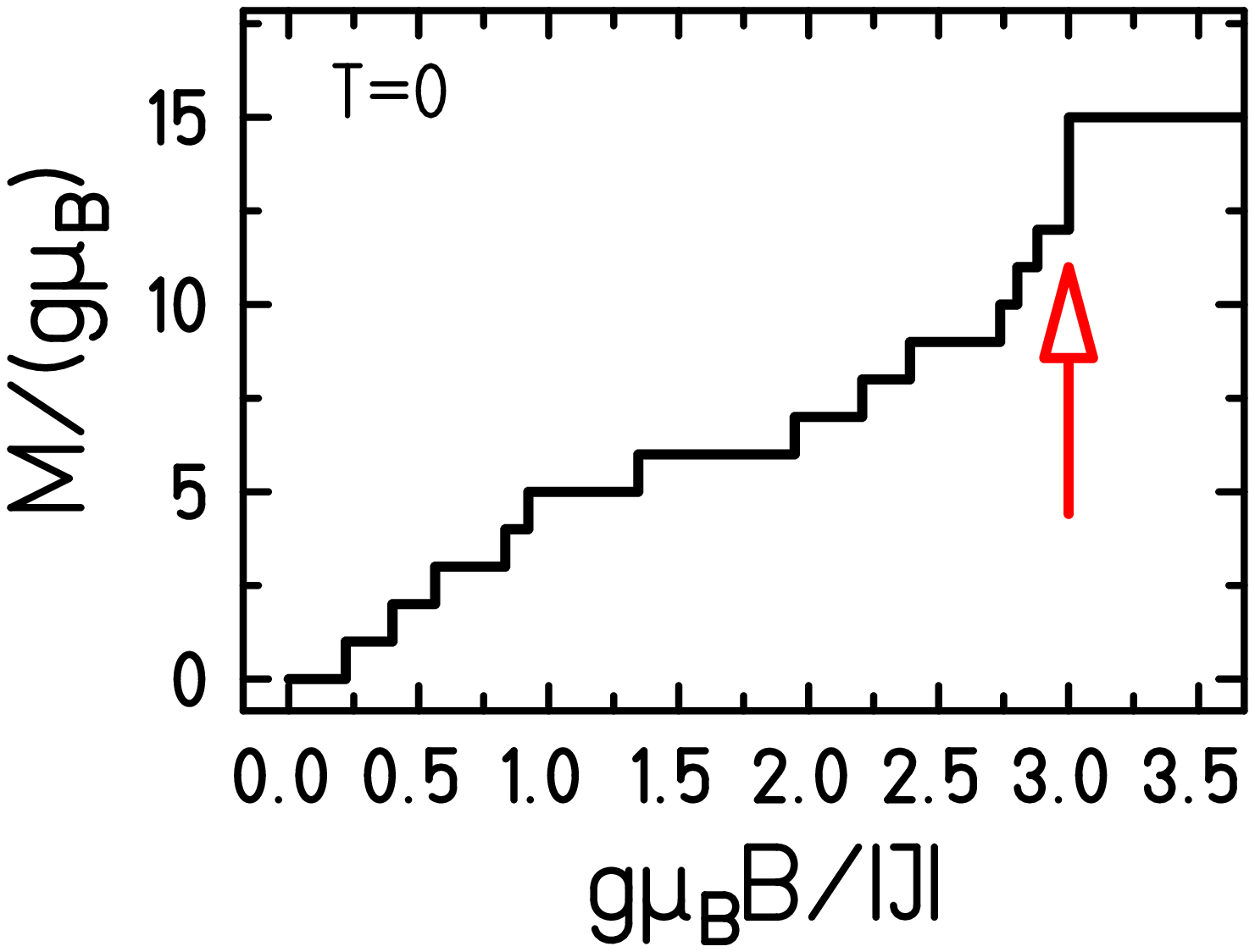}}
\caption{Top: Minimal energies of the icosidodecahedron for
  $s=1/2$. The highest four levels fall on a straight line.
  Bottom: Resulting $(T=0)$-magnetization curve. The
  magnetization jump of $\Delta M=3$ is marked by an arrow.}
\label{F-C}
\end{figure}

Nevertheless, the agreement of some properties with those of the
corresponding non-frustrated counterparts is only approximate.
Frustrated antiferromagnets exhibit their very own
characteristic features. One of them is given by independent
magnons which occur in certain spin arrays. It has been found
that the minimal energies $E_{\text{min}}(S)$ of
antiferromagnetic molecules of cuboctahedral and
icosidodecahedral structure depend linearly on total spin $S$
above a certain total spin \cite{SSR:EPJB01}.  Such a
dependence, which is depicted in \figref{F-C}(top) for an
icosidodecahedron with intrinsic spin $s=1/2$, results in an
unusually big jump to saturation as can be seen in \figref{F-C}
(bottom). Although first noticed for the Keplerate molecule
\mofe, such a behavior is quite common for a whole class of
frustrated spin systems including for instance the kagome or the
pyrochlore lattice antiferromagnet \cite{SHS:PRL02,RSH:JPCM03}
as well as for other frustrated magnetic molecules
\cite{SSR:JMMM05}. The underlying reason is that due to the
special geometric frustration in such systems -- even polygons
are surrounded by triangles -- the relative ground states in
subspaces ${\mathcal H}(M)$ are for big enough $M$ given by
product states of independent localized magnons.  This is one of
the rare examples where eigenstates of an interacting many-body
Hamiltonian can be constructed analytically.

\begin{figure}[ht!]
\centering
\centerline{\phantom{xxxxx}\includegraphics[clip,width=50mm]{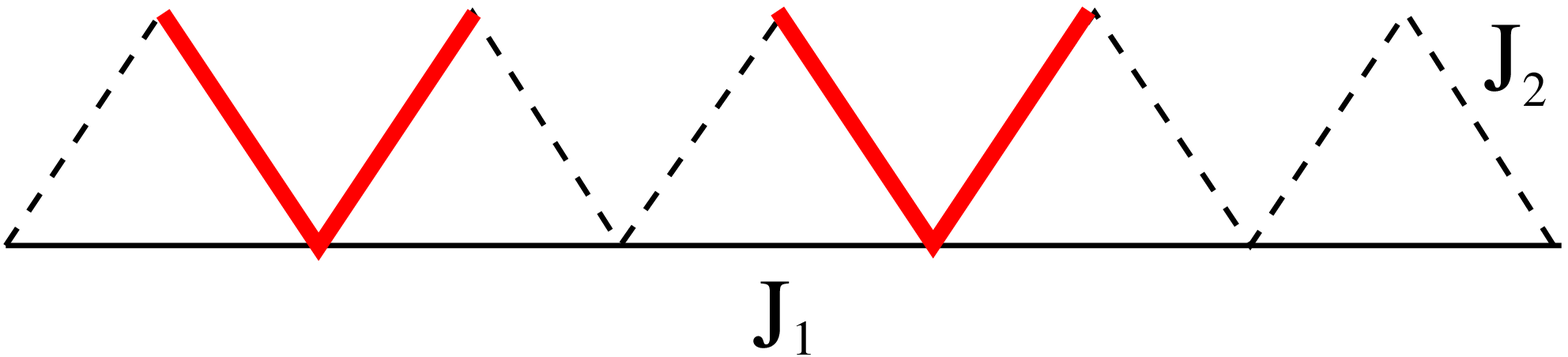}}

\vspace*{5mm}

\centerline{\includegraphics[clip,width=55mm]{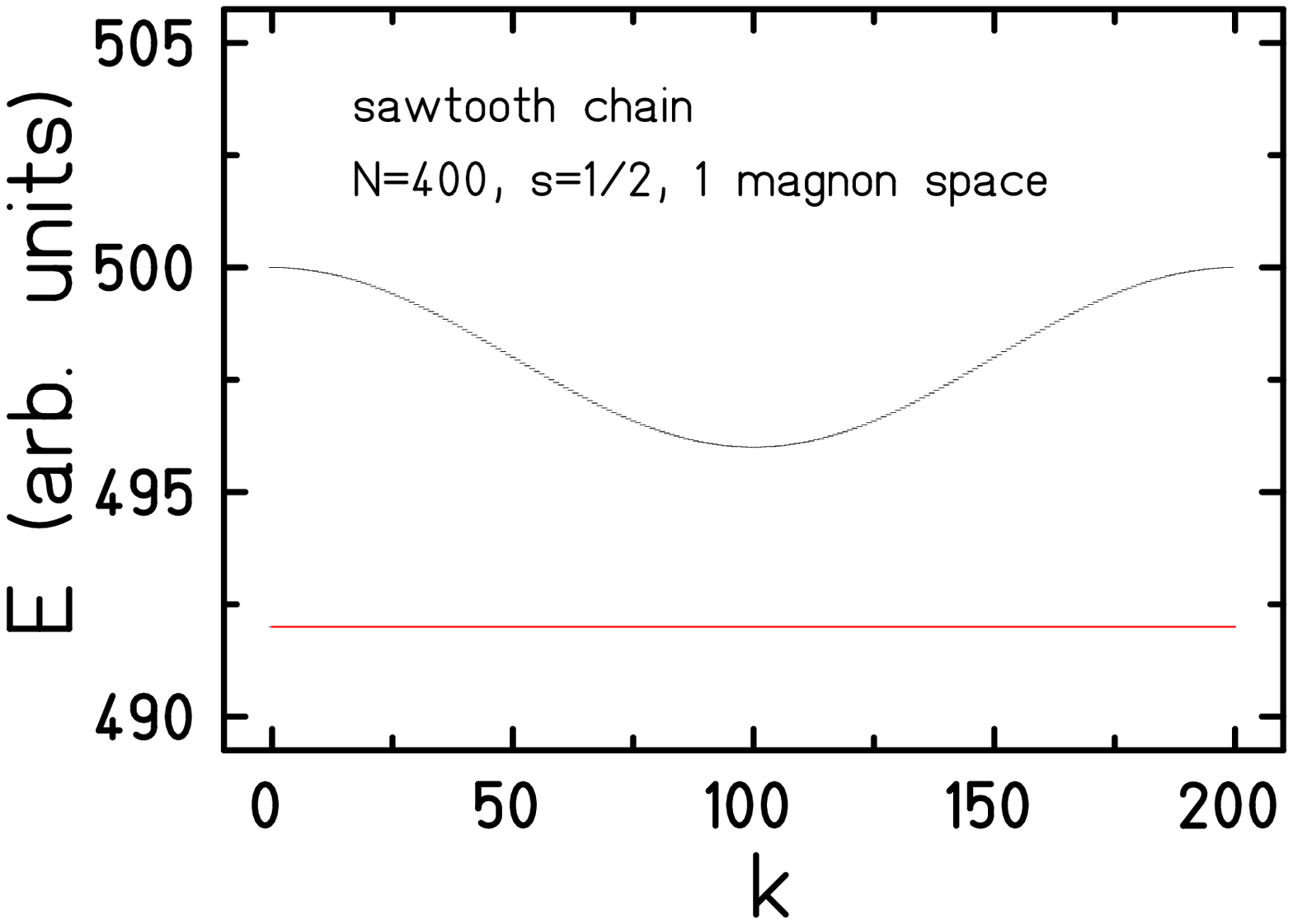}}
\caption{Top: Structure of the sawtooth chain, which hosts
  independent magnons (highlighted) for $J_2=2 J_1$.
  Bottom: Energy bands in one-magnon space; the flat band
  consists of $N/2$ degenerate levels.}
\label{F-Z}
\end{figure}

From a condensed matter point of view the occurrence of
independent and localized magnons can be understood by looking
at the band structure in one-magnon space. Figure~\xref{F-Z}
shows as an example the energy levels of the sawtooth chain, a
structure that can exhibit a truly giant magnetization jump of
half of the saturation magnetization \cite{RSH:JPCM03}. In
one-magnon space the energy levels form two bands: an
$N/2$-times degenerate (flat) ground state band and a cosine
shaped excited band. It is clear that the degenerate states of
the flat band can be superimposed in order to form localized
objects which are independent if well-enough separated. The
occurrence of flat bands seems to be rather common
\cite{MKO:JPSJ05}, they exist for instance in the afore discussed
magnetic molecules of icosidodecahedral and cuboctahedral
structure as well as in the kagome and the pyrochlore
antiferromagnet \cite{RSH:JPCM03}.

\begin{figure}[ht!]
\centering
\centerline{\includegraphics[clip,width=55mm]{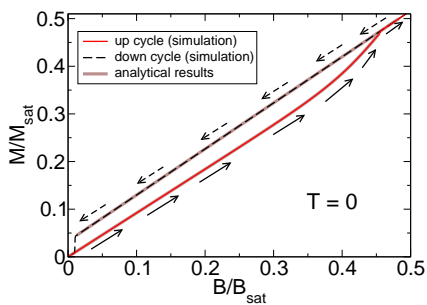}}
\caption{Hysteresis behavior of the classical icosahedron in an
  applied magnetic field obtained by classical spin dynamics
  simulations (thick lines) as well as by analytical stability
  analysis (grey lines). The curves match perfectly
  \cite{SSS:PRL05}.} 
\label{F-A}
\end{figure}

As the last point of this section I want to discuss the
observation that hysteresis can occur without anisotropy.
Usually the observation of hysteresis effects in magnetic
materials is an outcome of their magnetic anisotropy.  In a
recent article we could report that magnetic hysteresis
(\figref{F-A}) occurs in a spin system without any anisotropy
\cite{SSS:PRL05}.  Specifically, we investigated an icosahedron
where classical spins mounted on the vertices are coupled by
antiferromagnetic isotropic nearest-neighbor Heisenberg
interaction giving rise to geometric frustration.  At $T=0$ this
system undergoes a first order metamagnetic phase transition at
a critical field between two distinct families of ground state
configurations. The metastable phase of the system is
characterized by a temperature and field dependent survival
probability distribution.  Our exact classical treatment shows
that the abrupt transition at $T=0$ originates in the
intersection of two energy curves belonging to different
families of spin configurations that are ground states below and
above the critical field. The minimum of the two energy
functions constitutes a non-convex minimal energy function of
the spin system and this gives rise to a metamagnetic phase
transition \cite{LhM:02}.  We could also show that the
corresponding quantum spin system for sufficiently large spin
quantum number $s\ge 4$ possesses a non-convex set of lowest
energy levels when plotted versus total spin \cite{SSS:PRL05}
which again leads to a nontrivial magnetization jump.
Nevertheless, the investigation of the hysteretic behavior of
the quantum systems remains an open problem so far since the
magnetization dynamics should involve the modeling of the
relaxation of the quantum spin system in a time-dependent field
and coupled to a thermostat. This future investigation will
either be performed by solving the complete Schr\"odinger
dynamics of the coupled spin and bath systems or by following
effective Master equations, compare e.g. Ref.~\cite{HMH:PRE05}.

\section{Frustration effects of the three-leg ladder compound
  [(CuCl$_2$tachH)$_3$Cl]Cl$_2$}

\begin{figure}[ht!]
\centering
\centerline{\phantom{xxxxx}\includegraphics[clip,width=40mm]{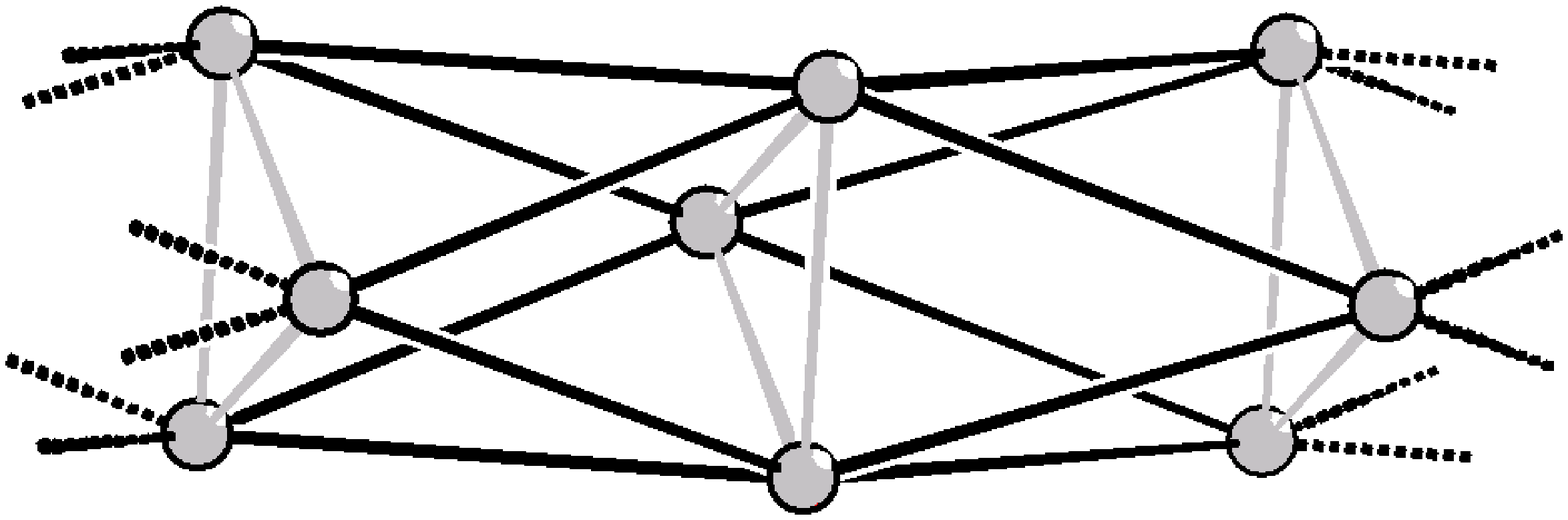}}

\vspace*{5mm}

\centerline{\includegraphics[clip,width=50mm]{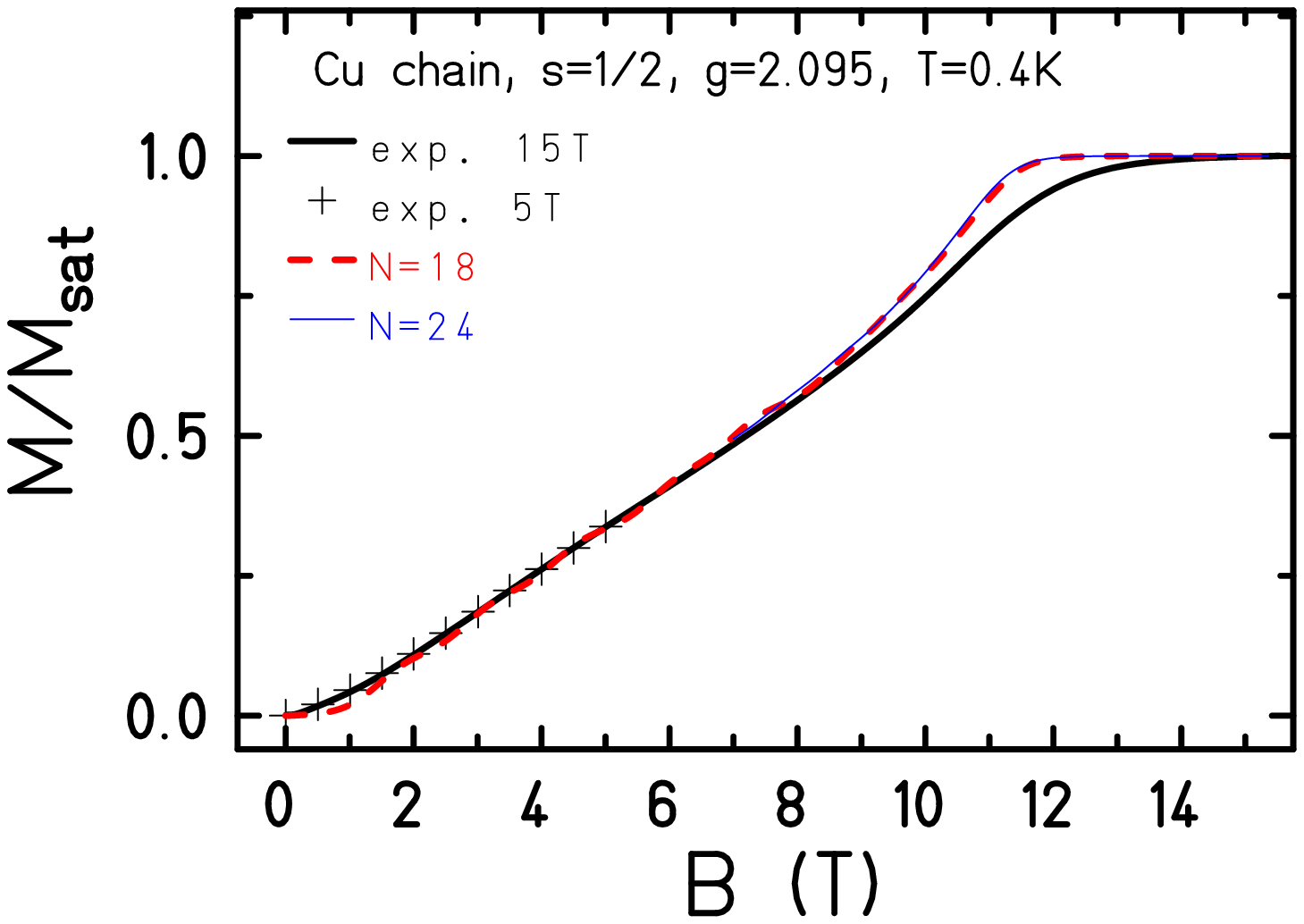}}
\caption{Top: Schematic structure of the triangular copper
  chain: The copper ions are placed at the vertices,
  intra-triangle exchange pathways are drawn by grey lines
  ($J_1=-0.9$~K), inter-triangle couplings are given by black lines
  ($J_2=-1.95$~K).  Bottom: Magnetization per copper triangle:
  Experimental data taken at a cryostat temperature of 0.4~K are
  given by thick solid lines, the two data sets are practically
  identical. The theoretical estimates are given for 18 and 24
  spins, i.~e. 6 or 8 triangles. The saturation field is
  $B_{\text{sat}}=11$~Tesla \cite{SNK:PRB04}.}
\label{F-Y}
\end{figure}

Recently we reported the magnetic features of a new
one-dimensional stack of antiferromagnetically coupled
equilateral copper(II) triangles, see top of \figref{F-Y}
\cite{SKK:CC04,SNK:PRB04}. High-field magnetization measurements
(bottom of \figref{F-Y}) showed that the interaction between the
copper triangles is of the same order of magnitude as the
intra-triangle exchange although only coupled via hydrogen
bonds. Thus, this molecule-based infinite chain turns out to be
an interesting example of a frustrated cylindrical three-leg
ladder with competing intra- and inter-triangle interactions.

\begin{figure}[ht!]
\centering
\centerline{\includegraphics[clip,width=50mm]{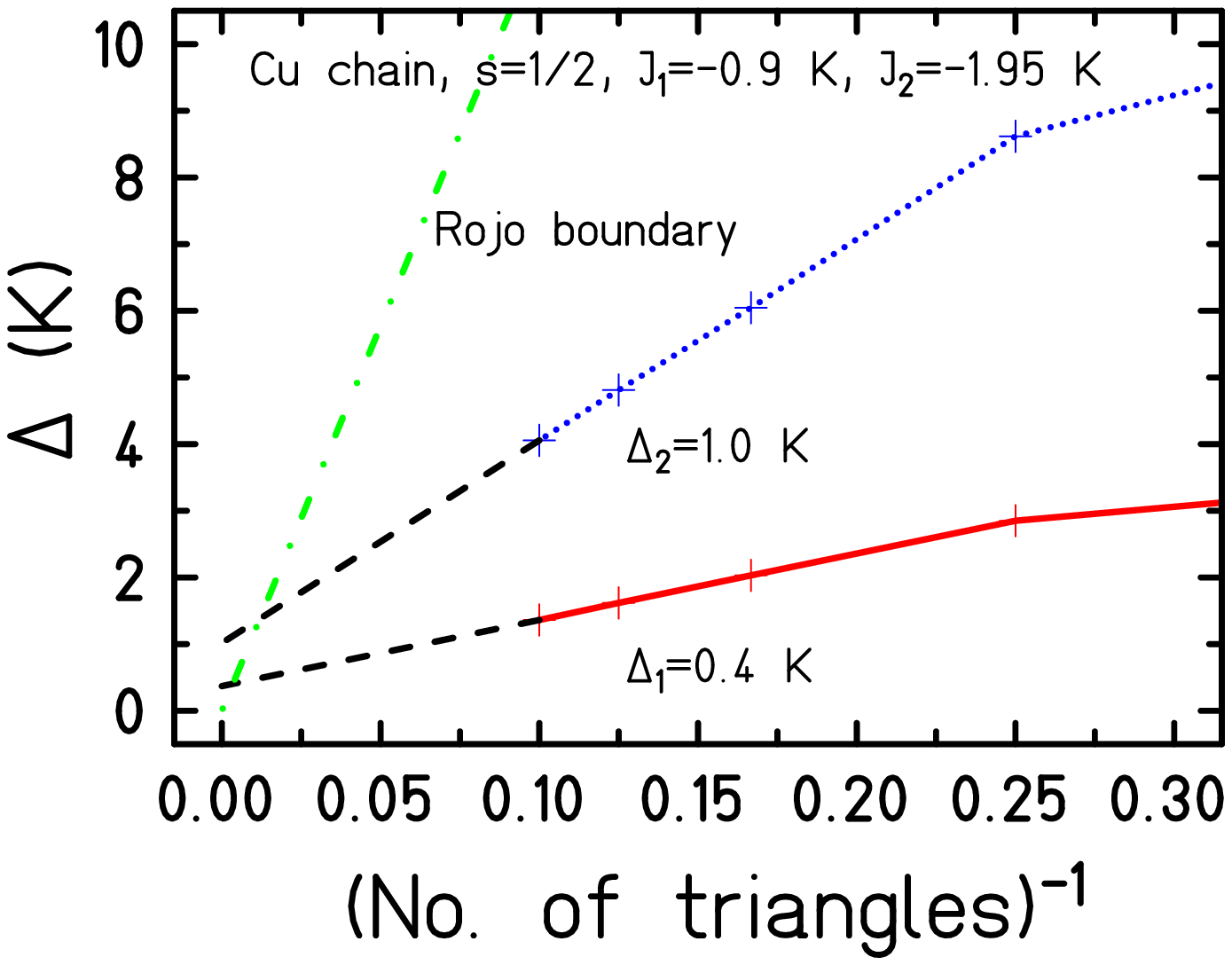}}

\vspace*{5mm}

\centerline{\includegraphics[clip,width=50mm]{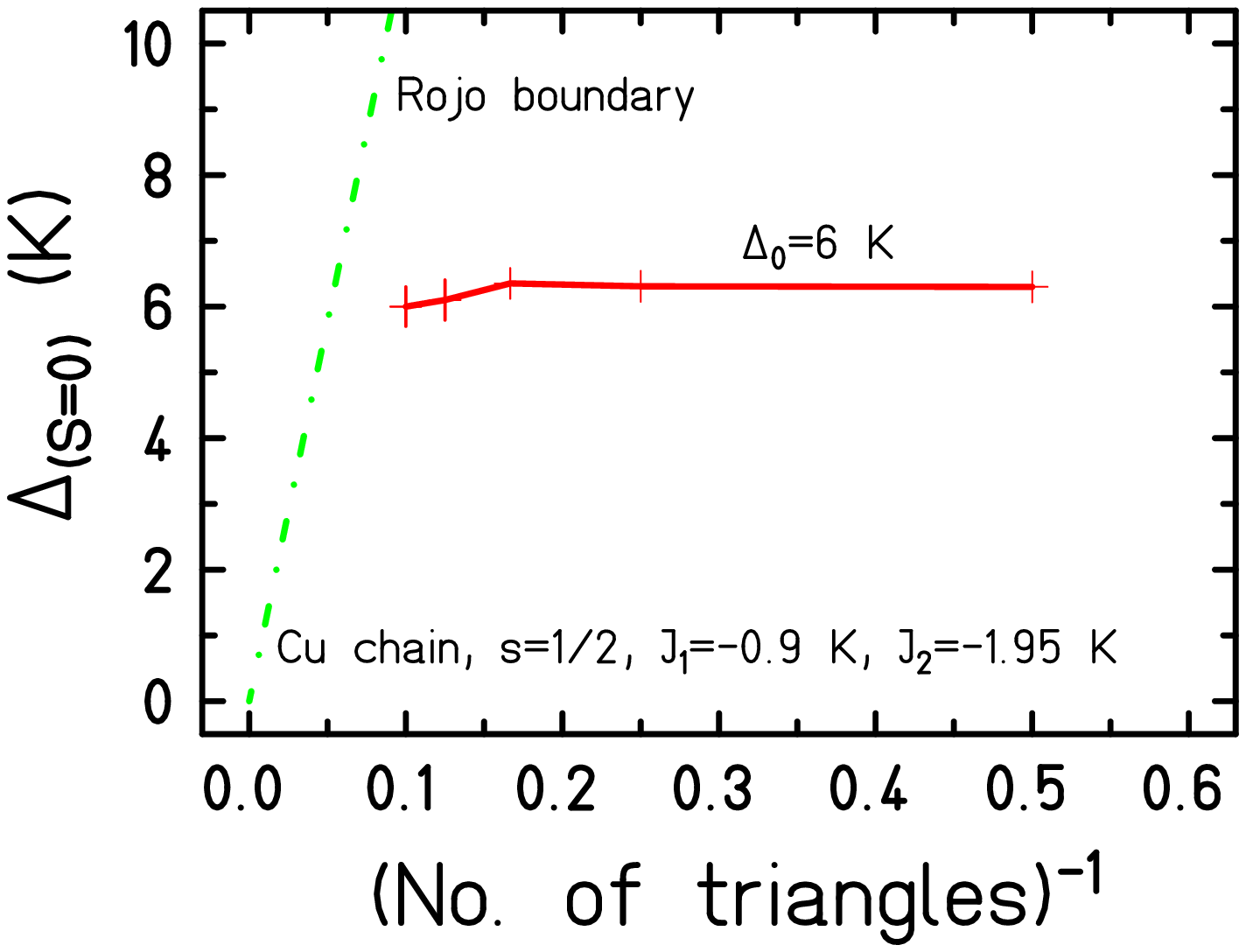}}
\caption{Top: Finite size extrapolation of the singlet-triplet
  gap $\Delta_1=E_{\text{min}}(S=1)-E_{\text{min}}(S=0)$ and the pentuplet-single gap
  $\Delta_2=E_{\text{min}}(S=2)-E_{\text{min}}(S=0)$ for
  [(CuCl$_2$tachH)$_3$Cl]Cl$_2$. Both gaps for the infinite chain
  are estimated to be non-zero.  Bottom:
  Finite size behavior of the singlet-singlet gap for
  [(CuCl$_2$tachH)$_3$Cl]Cl$_2$. The two leftmost data points
  for eight and ten triangles have an uncertainty of about
  $\pm0.2$~K due to the restricted accuracy of the Lanczos
  procedure for excited states \cite{SNK:PRB04}.
  The dashed line in both figures shows the upper boundary of
  the gap size according to Ref.~\cite{Roj:PRB96}.}
\label{F-X}
\end{figure}

A key question concerns the structure of the low-lying levels
and especially the existence of a singlet-triplet and a
singlet-singlet gap. Using finite size extrapolations of Lanczos
diagonalization results we demonstrated that the ground state is
a spin singlet which is gapped from the triplet excitation,
compare top panel of \figref{F-X}. It turned out that also the
first excited singlet is gaped from the ground state singlet,
see bottom panel of \figref{F-X}.

Our findings provoked theoretical investigations of related
models which yield that this frustrated spin system should not
have a gap \cite{LNM:PRB04,FLP:05}. The arguments are as follows
\begin{enumerate}
\item It is shown in Ref.~\cite{LNM:PRB04} that the Hamiltonian
  of a ladder compound such as [(CuCl$_2$tachH)$_3$Cl]Cl$_2$ can
  be mapped onto a simpler Hamiltonian of a linear chain of
  effective spins. For weak inter-triangle couplings $J_2$ each
  triangle basically remains in the triangular ground state with
  $S_{\text{triangle}}=1/2$, and thus the ladder should be an
  effective $s=1/2$ chain, which is augmented by additional
  chiral degrees of freedom~\cite{LNM:PRB04}. Such a chain
  should have a twofold degenerate ground state. In the opposite
  case where the inter-triangle coupling dominates, the
  triangular ground state should have $S_{\text{triangle}}=3/2$,
  and then the ladder should be an effective $s=3/2$ chain,
  which, since being a half-integer chain, is gapless.
\item Ref.~\cite{FLP:05} also provides a numerical argument. The
  Hamiltonian of [(CuCl$_2$tachH)$_3$Cl]Cl$_2$ is
  diagonalized for the relevant exchange parameters
  $J_1=-0.9$~K, $J_2=-1.95$~K with the help of Density Matrix
  Renormalization Group Techniques (DMRG). Using DMRG one can
  determine the ground state energies in subspaces ${\mathcal
    H}(M)$ of total magnetic quantum number $M$ for much larger
  chains. Nevertheless, due to technical problems with the
  boundary conditions which are intrinsic to the method only the
  gap to the pentuplet $E_{\text{min}}(S=2)-E_{\text{min}}(S=0)$
  could be determined, which indeed tends to zero.
\item The most general argument of Ref.~\cite{FLP:05} states
  that the three-leg ladder should either have a degenerate
  ground state or be gapless according to a proof given in
  Ref.~\cite{Roj:PRB96}. This proof is applicable for arbitrary
  ladder compounds with an odd number of sites on the rungs and
  half-integer intrinsic spin quantum number as long as the
  system possesses translational symmetry along the ladder. To
  the best of our knowledge this proof should be correct.
\end{enumerate}
The above arguments suggest that the finite size extrapolation
of Ref.~\cite{SNK:PRB04} might be biased by too short sample
systems (up to ten triangles). In addition the role of the
boundary conditions might be crucial for the actual numerical
convergence. In order to estimate, to which system size a
precise gap estimation must be performed in order to definitely
decide about low-lying gaps, the upper bound according to
Ref.~\cite{Roj:PRB96} is displayed as a dashed line in
\figref{F-X}. It seems realistic to reach conclusion about the
singlet-singlet gap, whereas the two other gaps
$\Delta_1=E_{\text{min}}(S=1)-E_{\text{min}}(S=0)$ and
$\Delta_2=E_{\text{min}}(S=2)-E_{\text{min}}(S=0)$ approach the
theoretical bound for rather large system sizes only.

Summarizing, the existence of low-lying gaps in the new
frustrated three-lag ladder appears to be ruled out for the
perfect model system. Investigations on the real spin system,
both theoretically and experimentally, will continue.

\section*{Acknowledgement}

It is my pleasure to thank K.~B\"arwinkel, A.~Honecker,
P.~K\"ogerler, M.~Luban, H.~Nojiri, J.~Richter, H.-J.~Schmidt,
and C.~Schr\"oder for the fruitful collaboration that produced
so many exciting results.  I am especially thankful to
H.-J.~Schmidt for valuable discussions about the three-lag
ladder compound. This work was supported by the Ph.D.  program
of the University of Osnabr\"uck as well as by the Deutsche
Forschungsgemeinschaft (Grant No. SCHN~615/5-1, SCHN~615/5-2,
and SCHN~615/8-1).


\end{document}